\documentclass[pra,twocolumn,amsmath,amssymb,aps,longbibliography,showpacs]{revtex4-1}
\usepackage{graphicx}
\usepackage{dcolumn}
\usepackage{bm}
\usepackage{CJK}
\usepackage{subfigure}
\usepackage{color}
\usepackage{epstopdf}
\newenvironment{sequation}{\small\begin{equation}}{\end{equation}}

\begin{document}


\title{Universal dynamics of superradiant phase transition in the anisotropic quantum Rabi model}

\author{Xunda Jiang$^{1,2}$}
\author{Bo Lu$^{1}$}
\author{Chengyin Han$^{1}$}
\author{Ruihuan Fang$^{1,2}$}
\author{Minhua Zhao$^{1,2}$}
\author{Zhu Ma$^{1,2}$}
\author{Tian Guo$^{1,2}$}
\author{Chaohong Lee$^{1,2,}$}
\email{lichaoh2@mail.sysu.edu.cn; chleecn@gmail.com}
\address{$^1$Guangdong Provincial Key Laboratory of Quantum Metrology and Sensing $\&$ School of Physics and Astronomy, Sun Yat-Sen University (Zhuhai Campus), Zhuhai 519082, China}
\address{$^2$State Key Laboratory of Optoelectronic Materials and Technologies, Sun Yat-Sen University (Guangzhou Campus), Guangzhou 510275, China}


\begin{abstract}
  We investigate the universally non-equilibrium dynamics of superradiant phase transition in the anisotropic quantum Rabi model.
  By introducing position and momentum operators, we obtain the ground states and their excitation gaps for both normal and superradiant phases via perturbation theory.
  We analytically extract the critical exponents from the excitation gap and the diverging length scale near the critical point, and find that the critical exponents are independent upon the anisotropy ratio.
  Moreover, by simulating the real-time dynamics across the critical point, we numerically extract the critical exponents from the phase transition delay and the diverging length scale, which are well consistent with the analytical ones.
  Our study provides a dynamic way to explore universal critical behaviors in the quantum Rabi model.
\end{abstract}

\maketitle


\section{INTRODUCTION}

Spontaneous symmetry breaking and quantum phase transitions (QPTs) are two fundamental and important concepts in physics.
The second-order QPTs always associate with spontaneous symmetry breaking~\cite{Sachdev2011,Morikawa1995,Kibble1980}, in which gapless energy spectra and degenerate ground states appear in the thermodynamical limit.
Due to the gapless excitations at the critical point, the adiabaticity breaks down when a system goes through a continuous phase transition.
As a consequence, nontrivial excitations such as domains~\cite{Kibble1976,Lee2009,Davis2011,Davis2012,Swislocki2013,Hofmann2014,Wu2017,Xu2016,Navon2015,Ye2018,Ye2018,Jiang2019}, vortices~\cite{Anderson2008,Su2013,Wu2016} and solitons~\cite{Damski2010,Witkowska2011,Zurek2009} appear spontaneously and obey the well-konwn Kibble-Zurek mechanism (KZM)~\cite{Kibble1976,Kibble1980,Zurek1985,Zurek1996,JDziarmaga2000,Polkovnikov2011,Bloch2008}.
The KZM has been extensively studied in various systems, from the early universe~\cite{Kibble1976,Kibble1980}, condensed matter systems~\cite{Ruutu1996,Bauerle1996,Monaco2009}, trapped ions~\cite{Campo2010,Ulm2013,Pyka2013,Ejtemaee2013,Lv2018}, to ultracold atomic gases~\cite{Anderson2008,Zurek2009,Lee2009,Damski2010,Witkowska2011,Davis2011,Davis2012,Navon2015,Lamporesi2013,Anquez2016,Clark2016,Feng2018,Swislocki2013,Hofmann2014,Wu2017,Xu2016,Ye2018,Ye2018,Jiang2019}.

The quantum Rabi model (QRM), a paradigmatic model in quantum optics, describes the fundamental interaction between quantized fields and two-level quantum systems~\cite{Forn2019,Kockum2019,Rabi1936,Rabi1937,Zhong2013,Zhong2014,Xie2017}.
In the thermodynamic limit, the QRM exhibits normal-superradiant phase transition, which provides an excellent platform for exploring universal behavior in both equilibrium~\cite{Ashhab2013,Bishop1996,Larson2017,Hwang2010} and non-equilibrium dynamics~\cite{Hwang2018,Puebla2017,Hwang2015}.
The anisotropic QRM, whose rotating and counter-rotating interactions have different coupling strengths~\cite{Fan2014,Zhang2017,WangMY2018}, is a generalized QRM.
In recent, QPTs in the anisotropic QRM and their universality are studied~\cite{Lin2017}.
However, the corresponding non-equilibrium universal dynamics is still unclear, it is worthy to clarify whether the anisotropic ratio affects the universality.

In this work, we investigated the non-equilibrium universal dynamics in the anisotropic QRM.
Under the description of position and momentum operators, making use of the Schrieffer-Wolff (SW) transformation, we obtain the ground states and their excitation gaps with the second-order perturbation theory.
Then, we analytically extract the critical exponents from the excitation gap and the diverging length scale, which reveal that the anisotropic QRM shares the same critical exponents for different anisotropy ratios between the rotating and counter-rotating terms.
Furthermore, we numerically simulate the real-time dynamics of the anisotropic QRM whose coupling strength is linearly swept across the critical point.
With the non-equilibrium dynamics, we numerically extract two universal scalings from the phase transition delay and the diverging length scale with respect to the quench time.
The critical exponents extracted from the numerical simulation are well consistent with the analytical ones.

The paper is organized as follows.
In Sec.~\uppercase\expandafter{\romannumeral2}, we introduce the  anisotropic QRM and give its ground states and excitation gaps.
In Sec.~\uppercase\expandafter{\romannumeral3}, we briefly review the KZM, and analytically extract the critical exponents from the excitation gap and the variance of the position and momentum operators.
In Sec.~\uppercase\expandafter{\romannumeral4}, we present the real-time non-equilibrium universal dynamics, and extract the critical exponents from the phase transition delay and the diverging length scale.
Finally, we give a brief summary and discussion in Sec.~\uppercase\expandafter{\romannumeral5}.

\section{The Anisotropic Quantum Rabi Model: Ground states and Excitation gaps}

In the units of $\hbar=1$, the anisotropic QRM can be described by the full-quantum Hamiltonian,
\begin{equation}
H = \omega {a^\dag }a + \frac{\Omega }{2}{\sigma _x} + g\left[ {\left( {{\sigma _ + }a + {\sigma _ - }{a^\dag }} \right) + \lambda \left( {{\sigma _ + }{a^\dag } + {\sigma _ - }a} \right)} \right],     \label{Hamiltonian0}
\end{equation}
where $a^{\dagger}(a)$ are the creation (annihilation) operators of the phonons with frequency $\omega$, $g$ is the coupling strength and $\lambda$ denotes the anisotropic ratio between rotating and counter-rotating terms.
Given the Pauli matrices $\sigma_{x,y,z}$, the second term describes a two-level system ${\sigma _ \pm } = \left( {{\sigma _z} \mp i{\sigma _y}} \right)/2$ with a transition frequency $\Omega$.

\begin{figure}[htbp]
\centering
\includegraphics[scale=0.43]{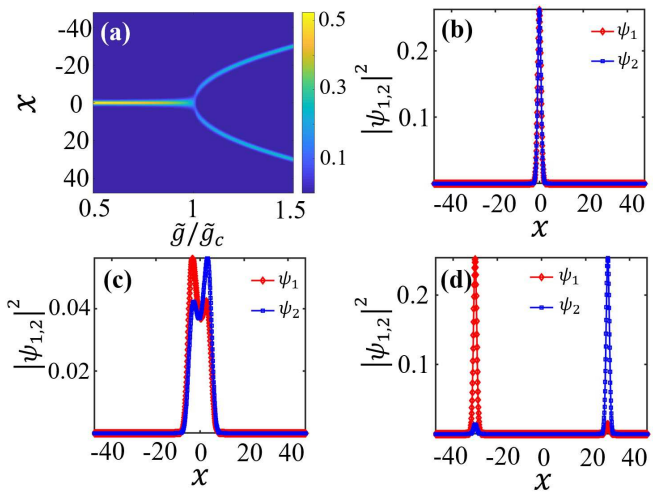}
\caption{(Color online) Density distributions of the ground states of the quantum Rabi model. (a) The total density distribution $\left|\psi_{1}\right|^{2}+\left|\psi_{2}\right|^{2}$ for $\lambda=1$, $\omega=1$, $\Omega=1000$ and $L=96$; (b,c,d) are three typical density distribution for $\tilde{g}/\tilde{g}_{c}=0.5$, $1.02$ and $1.5$, respectively. In which, $\tilde{g}/\tilde{g}_{c}=1$ is the critical point.}
\label{GS_1}
\end{figure}

Defining the dimensionless position and momentum operators $x = \left( {a + {a^\dag }} \right){\rm{   }}/{\sqrt 2 }$ and $p ={i}\left( {{a^\dag } - a} \right)/{\sqrt 2 }$, the Hamiltonian reads
\begin{small}
\begin{equation} \label{Hamiltonian1}
H = \frac{\omega }{2}\left( {{p^2} + {x^2}} \right) + \frac{\Omega }{2}{\sigma _x} + \tilde g\sqrt {\frac{{\Omega \omega }}{8}} \left[ {\left( {1 + \lambda } \right){\sigma _z}x + \left( {1 - \lambda } \right){\sigma _y}p} \right],
\end{equation}
\end{small}
where $\tilde{g}=2g/\sqrt{\Omega\omega}$.
The Hamiltonian becomes the QRM when $\lambda=1$, while it is the JC model when $\lambda=0$.
The second term becomes dominant in the limit $\Omega/\omega \to \infty$, thus the relevant low-energy states have $\langle \sigma_{x} \rangle \simeq -1$.
Within this subspace, the ground states can be determined by the competition between the first term (a conventional oscillator) and the last term (the coupling between the phonon field and the two-level system)~\cite{Luo2015,Lin2017}.
In Fig.~\ref{GS_1}, we show the typical ground state of Eq.~(\ref{Hamiltonian1}) for different coupling strength.
It clearly show the ground states undergoes a spontaneous symmetry breaking from symmetric to asymmetric when $\tilde{g}$ increases, see Fig.~\ref{GS_1}(a).

In the weak coupling region, $\tilde{g}<\tilde{g}_{c}$, the ground state is dominated by the oscillator term in the  normal phase as shown in Fig.~\ref{GS_1}(b), which is the vacuum of the phonon field and atom in the low energy space.
However, as the field-matter coupling is increased to the deep strong coupling regime $\tilde{g}>\tilde{g}_{c}$, the ground state turns from the normal phase to the superradiant phase, in which both the atom and the phonon field become excited [see Fig.~\ref{GS_1}(c,d)].
To describe the superradiant phase transition, one may choose the excitation of the atom and the phonon field can be served as an order parameter~\cite{Larson2017,Hwang2015}.

\subsection{Normal phase}

Below we briefly review the derivation of ground states and their excitation gaps from a low-energy effective Hamiltonian~\cite{Lin2017}.
To obtain the low-energy effective Hamiltonian, one may apply the SW transformation.
The Hamiltonian~(\ref{Hamiltonian1}) includes an unperturbed Hamiltonian $H_{0}$ and a off-diagonal perturbation $H_{V}$ as follows
\begin{equation}
\begin{array}{l}
{H_0} = \frac{\omega }{2}\left( {{p^2} + {x^2}} \right) + \frac{\Omega }{2}{\sigma _x}, \\
{H_V} = \tilde g\sqrt {\frac{{\Omega \omega }}{8}} \left[ {\left( {1 + \lambda } \right){\sigma _z}x + \left( {1 - \lambda } \right){\sigma _y}p} \right].
\end{array}
\end{equation}
By introducing a unitary operator $S_{0}$~\cite{Lin2017}, which is the generator of the SW transformation and is an anti-Hermitian operator,
\begin{equation}
  {S_{0}}  = i\tilde g\sqrt {\frac{\omega }{{8\Omega }}} \left[ {\left( {1 - \lambda } \right){\sigma _z}p - \left( {1 + \lambda } \right){\sigma _y}x} \right].
\end{equation}
Therefore the second-order low-energy effective Hamiltonian is given as
\begin{equation}\label{Eff_Ham1}
H_{eff}^{(2)} = \left\langle  -  \right|H_{eff}^{\left( 2 \right)}\left|  -  \right\rangle  \simeq \frac{\omega }{2}\left( {1 - {\xi ^{'}}^2} \right){p^2} + \frac{\omega }{2}\left( {1 - {\xi ^2}} \right){x^2},
\end{equation}
where $\left| \pm \right\rangle$ are the eigenstates of $\sigma_{x}$, $ \xi  = \tilde g\left( {1 + \lambda } \right)/2$ and ${\xi ^{'}} =  \tilde g\left( {1 - \lambda } \right)/2$.
In the weak coupling region,  the effective Hamiltonian behaves as the conventional harmonic oscillator as shown in Fig.~\ref{GS_1}(b), which corresponds to the normal phase.
The normal phase is the vacuum of atom and phonon excitations.
The excitation gap is given as
\begin{equation}\label{Energy_Gap_0}
  \varpi_{0} =\omega \sqrt {\left( {1 - {\xi ^2}} \right)\left( {1 - {\xi ^{'}}^2} \right)}.
\end{equation}
The normal-to-superradiant phase transition occurs at $\varpi _{0}=0$, which gives $\xi_{c}=1$ or $\xi^{'}_{c}=1$, that is,
\begin{equation}
{{\tilde g}_c} = \frac{2}{{1 + \left| \lambda  \right|}}.
\end{equation}
In the region of $\tilde{g} \le \tilde{g}_{c}$, the ground state is a normal phase ${\psi _0}\left( x,\alpha_{0} \right) = {e^{ - {S_{0}}}}{\phi _0}\left( x,\alpha_{0} \right)\left|  -  \right\rangle$ with
\begin{equation}\label{Ground_state}
  {\phi _0}\left( x,\alpha_{0} \right) = \frac{{\sqrt \alpha_{0}  }}{{{\pi ^{1/4}}}}\exp \left( { - \frac{1}{2}{\alpha_{0} ^2}{x^2}} \right)
\end{equation}
denoting the ground state of the harmonic oscillator.
Here the effective mass $m_{0} = 1/\left[\omega\left( {1 - {\xi ^{'}}^2} \right)\right]$ and the wavepacket width $\alpha_{0}  = \sqrt{m_{0} \varpi_{0}}$.

\subsection{Superradiant phase}

We now discuss the ground states and the corresponding excitation gaps for the superradiant phase.
In the region of $\tilde{g}>\tilde{g}_{c}$, the system enters into the  superradiant phase and the effective Hamiltonian~(\ref{Eff_Ham1}) for the normal phase breaks down.
This means that $P = \left| - \right\rangle \left\langle - \right|$ is no longer the suitable low-energy subspace.
Making use of the SW transformation, we introduce new generators to diagonalize the Hamiltonian for both $\lambda>0$ and $\lambda<0$.
Then, one may obtain an effective Hamiltonian and give its ground states and excitation gaps.

\begin{figure}[htb]
\centering
\includegraphics[scale=0.42]{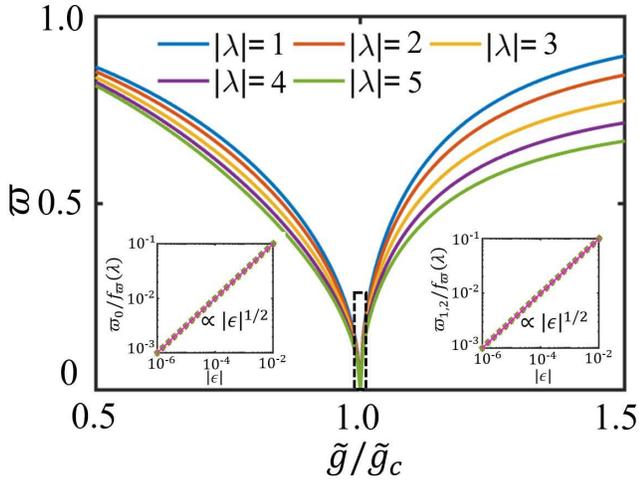}
\caption{(Color online) The excitation gap obtained from the second-order perturbative Hamiltonian.
Insets: the universal scalings of the energy gaps near the critical region labelled by the dashed-line rectangle, where the left and right insets respectively correspond to normal and superradiant phases.}
\label{Energy_Gap_Two_Phase_1}
\end{figure}

In the case of $\lambda>0$, we introduce a new displaced operator ${\cal D}_{1}\left[ {{\alpha }} \right] = {e^{ - i{\alpha}p}}=e^{-\alpha\frac{\partial}{\partial x}}$ with the parameter $\alpha$ to be determined, thus the Hamiltonian~(\ref{Hamiltonian1}) is transformed as
\begin{equation} \label{Dis_Ham_1}
\begin{array}{l}
H\left( \alpha  \right) = {{\cal D}_{1}^\dag }\left( \alpha  \right)H{\cal D}_{1}\left( \alpha  \right) = \frac{\omega }{2}\left( {{p^2} + {x^2}} \right) + \frac{\Omega }{2}{\sigma _x}\\
 + \sqrt {\frac{{\Omega \omega }}{2}} \left( {\xi {\sigma _z}x + {\xi ^{'}}{\sigma _y}p} \right) + \omega \alpha x + \frac{{\alpha \delta_{1} }}{2}{\sigma _z} + \frac{{\omega {\alpha ^2}}}{2},
\end{array}
\end{equation}
where $\delta_{1} ={\sqrt {2\Omega \omega } \xi }$.
The eigenstates of the atomic part $H_a=\frac{\Omega }{2}{\sigma _x} + \frac{{\alpha \delta_{1} }}{2}{\sigma _z}$ are
 \begin{equation}
   \left|  \uparrow  \right\rangle  = \cos \theta \left|  +  \right\rangle  + \sin \theta \left|  -  \right\rangle ,{\kern 6pt} \left|  \downarrow  \right\rangle  =  - \sin \theta \left|  +  \right\rangle  + \cos \theta \left|  -  \right\rangle,
 \end{equation}
with $\sin 2\theta  = \alpha \delta_{1} /\widetilde \Omega $, $ \cos 2\theta  = \Omega /\widetilde \Omega$ and the new atomic transition frequency $\widetilde \Omega  = \sqrt {{\Omega ^2} + {{\left( {\delta_{1} \alpha } \right)}^2}}$.
In terms of Pauli matrices $\tau_{x,y,z}$ associated with $\left\{ {\left|  \uparrow  \right\rangle ,\left|  \downarrow  \right\rangle } \right\}$, the Hamiltonian~(\ref{Dis_Ham_1}) becomes
\begin{equation}
  \begin{array}{l}
H\left( \alpha  \right) = \frac{\omega }{2}\left( {{p^2} + {x^2}} \right) + \sqrt {\frac{{\Omega \omega }}{2}} \left( {\xi \cos 2\theta x{\tau _x} - {\xi ^{'}}p{\tau _y}} \right)\\
{\kern 40pt}{\rm{ }} + \left( {\omega \alpha  + \sqrt {\frac{{\Omega \omega }}{2}} \sin 2\theta \xi {\tau _z}} \right)x + \frac{{\widetilde \Omega }}{2}{\tau _z} + \frac{{\omega {\alpha ^2}}}{2}.
\end{array}
\end{equation}
To eliminate the perturbation term, $\left( {\omega \alpha  + \sqrt {\Omega \omega /2} \sin 2\theta \xi {\tau _z}} \right)x$, we choose the parameter $\alpha$ such that $\omega \alpha  - \sqrt {\Omega \omega /2} \sin 2\theta \xi  = 0$,  which gives the nontrivial solutions
\begin{equation}
  \alpha  =  \pm {\alpha _g} =  \pm \sqrt {\left( {\Omega /2\omega {\xi ^2}} \right)\left( {{\xi ^4} - 1} \right)}.
\end{equation}
Given $\alpha  = \pm\alpha_{g}$, the Hamiltonian reads
\begin{equation}\label{Super_Ham_1}
  \widetilde H\left( { \pm {\alpha _g}} \right) = {\widetilde H_{\rm{0}}} + {\widetilde H_V},
\end{equation}
with
\begin{equation}
\begin{array}{l}
{\widetilde H_0} = \frac{\omega }{2}\left( {{p^2} + {x^2}} \right) + \frac{{\widetilde \Omega }}{2}{\tau _z},\\
{\widetilde H_V} = \sqrt {\frac{{\Omega \omega }}{2}} \left( {\xi \cos 2\theta x{\tau _x} - {\xi ^{'}}p{\tau _y}} \right).
\end{array}
\end{equation}
Making use of the SW transformation, we find a new generator,
\begin{equation}
{S_1} = i\sqrt {\frac{{\Omega \omega }}{{2{{\widetilde \Omega }^2}}}} \left( {{\xi ^{'}}p{\tau _x} + \xi \cos 2\theta x{\tau _y}} \right).
\end{equation}
for diagonalizing the Hamiltonian~(\ref{Super_Ham_1}) under the condition of $\tilde{\Omega}/\omega \gg 1$.
Thus the second-order low-energy effective Hamiltonian reads,
\begin{small}
\begin{equation}
\widetilde H_{eff}^{\left( 2 \right)} \simeq \frac{\omega }{2}\left( {1- \frac{\Omega }{{\widetilde \Omega }}{\xi ^{'}}^2} \right){p^2} + \frac{\omega }{2}\left( {1 - \frac{\Omega }{{\widetilde \Omega }}{\xi ^2}{{\cos }^2}2\theta } \right){x^2}.
\end{equation}
\end{small}
Comparing with the simple harmonic oscillator, the excitation gap (see Fig.~\ref{Energy_Gap_Two_Phase_1}) is given as
\begin{equation} \label{Energy_Gap_1}
  \varpi_{1} {\rm{ = }}\omega \sqrt {\left( {{\rm{1}} - \frac{1}{{{\xi ^4}}}} \right)\left( {{\rm{1}}{\kern 1pt} {\kern 1pt}  - \frac{{{\xi ^{'}}^{\rm{2}}}}{{{\xi ^2}}}} \right)}.
\end{equation}
The excitation gap recovers the previous result  when $\lambda=1$  ~\cite{Hwang2015}. Obviously, the $\varpi_{1}$ vanishes at $\xi_{c}=1$, which gives the critical point
\begin{equation}
 \tilde{g}_{c}=\frac{2}{1+\lambda}, (\lambda>0).
\end{equation}
The corresponding ground-state for $\tilde{g}>\tilde{g}_{c}$ is ${\psi _1}\left( x,\alpha_{1} \right) = {{\cal D}_1}\left( {{\alpha _g}} \right){e^{ - {S_1}}}{\phi _0}\left( {x,{\alpha _1}} \right)\left|  \downarrow  \right\rangle$, where $\phi_{0}\left( {x,{\alpha _1}} \right)$ is the ground state of the harmonic oscillator with $\alpha_{1}=\sqrt{m_{1}\varpi_{1}}$ and the effective mass $m_{1}=1/\left[{{\omega \left( {{\rm{1}} - {\xi ^{'}}^{\rm{2}}/{\xi ^2}} \right)}}\right]$.

In the case of $\lambda<0$, we introduce another displaced operator ${{\cal D}_2}\left( \beta  \right) = {e^{ - i\beta x}} = {e^{ - \beta \frac{\partial }{{\partial p}}}}$ with the parameter $\beta$ to be determined, thus the Hamiltonian~(\ref{Hamiltonian1}) is transformed as
\begin{equation}\label{Dis_Ham_2}
 \begin{array}{l}
H\left( \beta  \right) = {{\cal D}_2}^\dag \left( \beta  \right)H{{\cal D}_2}\left( \beta  \right) = \frac{\omega }{2}\left( {{p^2} + {x^2}} \right) + \frac{\Omega }{2}{\sigma _x}\\
 + \sqrt {\frac{{\Omega \omega }}{2}} \left( {\xi {\sigma _z}x + {\xi ^{'}}{\sigma _y}p} \right) + \omega \beta p + \frac{{\beta {\delta _2}}}{2}{\sigma _y} + \frac{{\omega {\beta ^2}}}{2},
\end{array}
\end{equation}
where $\delta_{2}={\sqrt {{\rm{2}}\Omega \omega } {\xi ^{'}}}$.
The eigenstates of the atomic part $H_a=\frac{\Omega }{2}{\sigma _x} + \frac{{\beta \delta_{2} }}{2}{\sigma _y}$ are
\begin{equation}
\left| {\widetilde  \uparrow } \right\rangle  = \cos {\theta ^{'}}\left|  +  \right\rangle  - i\sin {\theta ^{'}}\left|  -  \right\rangle ,{\kern 1pt}  \left| {\widetilde  \downarrow } \right\rangle  = \sin {\theta ^{'}}\left|  +  \right\rangle  + i\cos {\theta ^{'}}\left|  -  \right\rangle,
\end{equation}
with $\sin 2\theta^{'}  = \alpha \delta_{2} /\widetilde{\Omega}^{'}$, $ \cos 2\theta^{'} = \Omega /\widetilde{\Omega}^{'}$, and the new atomic transition frequency $\widetilde \Omega^{'}  = \sqrt {{\Omega ^2} + {{\left( {\delta_{2} \beta } \right)}^2}}$.
In terms of Pauli matrices $\tau^{'}_{x,y,z}$ associated with $\left\{ {\left|  \tilde{\uparrow}  \right\rangle ,\left|  \tilde{\downarrow}  \right\rangle } \right\}$, the new Hamiltonian reads
\begin{equation}
\begin{array}{l}
H\left( \beta  \right) = \frac{\omega }{2}\left( {{p^2} + {x^2}} \right) - \sqrt {\frac{{\Omega \omega }}{2}} \left( {\xi x{\tau^{'} _y} + {\xi ^{'}}\cos 2\theta p{\tau^{'} _x}} \right)\\
{\kern 35pt}  + \left( {\omega \beta  + \sqrt {\frac{{\Omega \omega }}{2}} \sin 2\theta {\xi ^{'}}{\tau^{'} _z}} \right)p + \frac{{\widetilde \Omega }}{2}{\tau^{'} _z} + \frac{{\omega {\beta ^2}}}{2}.
\end{array}
\end{equation}
To eliminate the perturbation term, $\left( {\omega \beta  + \sqrt {\Omega \omega /2} \sin 2\theta {\xi ^{'}}{\tau^{'} _z}} \right)p$, we choose $\omega \beta  - \sqrt {\Omega \omega /2} \sin 2\theta \xi^{'} = 0$, which gives
\begin{equation}
  \beta  =  \pm {\beta _g} = \sqrt {\left( {\Omega /2\omega {\xi ^{'}}^2} \right)\left( {{\xi ^{'}}^4 - 1} \right)}.
\end{equation}
Given $\beta=\pm\beta_{g}$, the Hamiltonian becomes
\begin{equation}\label{Super_Ham_2}
  \widetilde H^{'}\left( { \pm {\beta _g}} \right) = {\widetilde H_{\rm{0}}}^{'} + {\widetilde H_V}^{'},
\end{equation}
with
\begin{small}
\begin{equation}
\widetilde H_0^{'} = \frac{\omega }{2}\left( {{p^2} + {x^2}} \right) + \frac{{\widetilde \Omega }}{2}{\tau^{'} _z}, \widetilde H_V^{'} =  - \sqrt {\frac{{\Omega \omega }}{2}} \left( {\xi x{\tau^{'}_y} + {\xi ^{'}}\cos 2\theta p{\tau^{'}_x}} \right).
\end{equation}
\end{small}

Through performing SW transformation, under the condition of $\tilde{\Omega}^{'}/\omega \gg 1$, we diagonalize the Hamiltonian~(\ref{Super_Ham_2}) with the generator
\begin{equation}
S_{2}= i\sqrt {\frac{{\Omega \omega }}{{2{{\widetilde \Omega }^2}}}} \left( {\xi x{\tau^{'}_x} - {\xi ^{'}}\cos 2\theta p{\tau^{'}_y}}\right).
\end{equation}
Then we obtain the second-order low-energy effective Hamiltonian,
\begin{equation}
{\widetilde{H^{'}}}_{eff}^{(2)} \simeq \frac{\omega }{2}\left( {1 - \frac{\Omega }{{\widetilde \Omega }}{\xi ^{'2}}{{\cos }^2}2\theta } \right){p^2} + \frac{\omega }{2}\left( {1 - \frac{\Omega }{{\widetilde \Omega }}{\xi ^2}} \right){x^2},
\end{equation}
and the excitation gap (see Fig.~\ref{Energy_Gap_Two_Phase_1})
\begin{equation} \label{Energy_Gap_2}
\varpi_{2}{\rm{ = }}\omega \sqrt {\left( {1 - \frac{1}{{{\xi ^{'}}^4}}} \right)\left( {1 - \frac{{{\xi ^2}}}{{{\xi ^{'}}^2}}} \right)}.
\end{equation}
Obviously, the excitation gap $\varpi_{2}$ vanishes at the critical point $\xi^{'}_{c}=1$, that is,
\begin{equation}
 \tilde{g}_{c}=\frac{2}{1-\lambda}, (\lambda<0).
\end{equation}
The corresponding ground state is ${\psi _2}\left( x,\alpha_{2} \right) = {{\cal D}_2}\left( {{\beta _g}} \right){e^{ - {S_2}}}{\phi _0}\left( {x,{\alpha _2}} \right)\left|  \tilde{\downarrow}  \right\rangle$, where $\phi_{0}\left( {x,{\alpha _2}} \right)$ is the ground state of the simple harmonic oscillator  with $\alpha_{2}=\sqrt{m_{2}\varpi_{2}}$ and the effective mass $m_{2}=1/\left[\omega \left( {1 - 1/{\xi ^{'4}}} \right)\right]$.

\section{Universal critical dynamics across superradiant phase transition}

\subsection{Analytical Kibble-Zurek scalings}
In the following, we briefly introduce the KZM and analytically derive the universal critical exponents.
Near the quantum critical point, due to the vanishing of the energy gap,  the
correlation (or healing) length $\zeta$ and relaxation time $\tau$ diverge as
\begin{equation}\label{Healing_Relaxation_Scaling}
  \tau  \sim {\left| \epsilon \right|^{-vz}},{\kern 10pt}  \zeta  \sim {\left| \epsilon \right|^{ - v}},
\end{equation}
where  $\epsilon$ is the dimensionless distance from the critical point, and $(v,z)$ are the critical exponents.
To drive system from normal to superradiant phase,  we linearly quench the coupling strength  according to
\begin{equation}\label{Dim_parameter}
  \epsilon\left(t\right)=\frac{|\tilde{g}\left(t\right)-\tilde{g}_{c}|}{\tilde{g}_{c}}=\frac{t}{\tau_{Q}},
\end{equation}
where $\tau_{Q}$ is the quench time.  In a QPTs, the relaxation time is defined by the inverse of the
 gap between the ground state and the first relevant excited state, i.e. $ \tau  \simeq {\varpi ^{ - 1}}$.
However, the relaxation time is divergent in the vicinity of the critical point, in which the gap vanishes as
\begin{equation}\label{Energy_Scaling}
  \varpi \sim {\left| \epsilon \right|^{vz}}.
\end{equation}
When the transition rate $\left| \dot{\epsilon}/\epsilon\right|=1/\left|t\right|$ equals to the gap $\varpi \sim {\left| \epsilon \right|^{vz}}={\left| t/\tau_{Q} \right|^{vz}}$, the adiabaticity fails near an instant  $t=\hat{t}$ ,
\begin{equation}\label{Freeze_time}
  \hat{t}\sim \tau_{Q}^{\frac{vz}{1+vz}}, \quad   \hat{\epsilon}\sim\tau_{Q}^{-\frac{1}{1+vz}},
\end{equation}
and the corresponding correlation length becomes as
\begin{equation}
\hat{\zeta}\sim\hat{\epsilon}^{-v}\sim\tau_{Q}^{\frac{v}{1+vz}}.
\end{equation}

 In the region of $\tilde{g}\le \tilde{g}_{c}$,  the excitation gap $\varpi_{0}$  near the critical point vanishes as
\begin{equation} \label{Rabi_energy_scale}
\varpi_{0}\propto  f_{{\varpi }}\left( \lambda  \right){\left| \epsilon \right|^{1/2}}, {\kern 10pt} \lambda \neq0,
\end{equation}
where
\begin{equation}
  f_{{\varpi }}\left( \lambda  \right) = \omega \left[ {1 - {{\left( {\frac{{1 -\left|  \lambda \right| }}{{1 + \left| \lambda  \right| }}} \right)}^{\rm{2}}}} \right]^{1/2}.
\end{equation}
Comparing  with Eq.~(\ref{Energy_Scaling}), we analytically obtain $vz=1/2$.
When $\lambda=0$, the excitation gap becomes
\begin{equation}
 \varpi_{0}  \propto   \left| \epsilon \right|^{1}.
\end{equation}
Given $vz=1$, the critical exponents for $\lambda=0$ are belong to a different universality class~\cite{Hwang2016}, which will not be discussed below.
For the anisotropic QRM, the energy gap $\varpi_{0}$ near the critical point vanishes, see the left insets of Fig.~\ref{Energy_Gap_Two_Phase_1}, it  clearly reveals that the anisotropic QRM shares the same universal class.

To extract the critical exponents, we introduce the position variance $\Delta x$ and the momentum variance $\Delta p$.
In the normal phase, $\Delta x$ and $\Delta p$ are obtained via the ground state $\psi_{0}$ .
\begin{equation}
\Delta x =\left[\frac{1}{2}\left( {1 - \frac{\omega }{\Omega }\xi {\xi ^{'}}} \right)\sqrt {\frac{{ {1 - {\xi ^{'}}^2} }}{{{1 - {\xi ^2}} }}}  + \frac{{\omega {\xi ^{'}}^2}}{{2\Omega }}\right]^{\frac{1}{2}},
\end{equation}
\begin{equation}
  \Delta p = \left[\frac{1}{2}\left( {1 - \frac{\omega }{\Omega }{\xi ^{'}}\xi } \right)\sqrt {\frac{{ {1 - {\xi ^2}} }}{{ {1 - {\xi ^{'}}^2} }}}  + \frac{{\omega {\xi ^2}}}{{2\Omega }}\right]^{\frac{1}{2}}.
\end{equation}
 Near the neighborhood of the phase transition, the length scale $\Delta x$ behaves as
 \begin{equation}
   \begin{array}{l}
\Delta x \propto f\left( \lambda  \right){\left| \epsilon \right|^{ - 1/4}}, {\kern 10pt}\lambda  > 0,\\
\Delta x \propto {f^{ - 1}}\left( \lambda  \right){\left| \epsilon \right|^{1/4}}, {\kern 6pt}\lambda  < 0,
\end{array}
 \end{equation}
where
 \begin{equation}
f\left( \lambda  \right) = {\left[ {1 - {{\left( {\frac{{1 - \left| \lambda  \right|}}{{1 + \left| \lambda  \right|}}} \right)}^{\rm{2}}}} \right]^{1/4}}.
 \end{equation}
The critical behavior of $\Delta x$ shows that it is divergent when $\lambda>0$, while it vanishes when $\lambda<0$.
It's worthy to note that $\Delta x$ plays an analogous role of the diverging length scale when $\lambda>0$~\cite{Sachdev2011,Hwang2015}.
Comparing with  Eq.~(\ref{Healing_Relaxation_Scaling}), we obtain the static correlation length critical exponent $v=1/4$ and the dynamic critical exponent $z=2$.

For the momentum variance $\Delta p$, its critical behavior obeys
\begin{equation}
  \begin{array}{l}
\Delta p \propto {f^{ - 1}}\left( \lambda  \right){\left| \epsilon \right|^{1/4}},{\kern 7pt}  \lambda  > 0,\\
\Delta p \propto f\left( \lambda  \right){\left| \epsilon \right|^{ - 1/4}},{\kern 12pt}  \lambda  < 0.
\end{array}
\end{equation}
In contrast to $\Delta x$,  $\Delta p$ becomes divergent when $\lambda<0$, while it vanishes when $\lambda>0$.
In the case of $\lambda<0$, the diverging length scale $\Delta p$ gives the critical exponent $v=1/4$ according to KZM, and so that we have the critical exponent $z=2$.

In the region of $\tilde{g} > \tilde{g}_{c}$,  we divide the superradiant phase into two parts, which label as $x$-type($p$-type) superradiant phase when $\lambda > 0 (\lambda < 0)$~\cite{Lin2017}, respectively.
In the superradiant phase, the excitation gap $\varpi_{1,2}$ near the critical point vanishes as
 \begin{equation}
\varpi_{1,2} \propto  f_{{\varpi }}\left( \lambda  \right){\left| \epsilon \right|^{1/2}},
{\kern 10pt} \lambda \neq0.
 \end{equation}
The critical behaviors  of the excitation gap are shown in the right insets of Fig.~\ref{Energy_Gap_Two_Phase_1}, which  gives $vz=1/2$.
In the $x$-type superradiant phase,  $\Delta{x}$ and $\Delta{p}$  are obtained via
the ground state ${\psi _1}\left( x,\alpha_{1} \right) $.
\begin{small}
\begin{equation}
 \Delta x = {\left[ {\frac{1}{2}\left( {1 - \frac{{\omega {\xi ^{'}}}}{{\Omega {\xi ^5}}}} \right)\sqrt {\frac{{{\xi ^2} - {\xi ^{'2}}}}{{{\xi ^2} - {\xi ^{ - 2}}}}}  + \frac{{\omega {\xi ^{'2}}}}{{2\Omega {\xi ^4}}} - {\frac{{{\xi ^{'}}}}{{2{\xi ^3}}} + \frac{{{\xi ^{'}}}}{{2{\xi ^7}}}} } \right]^{\frac{1}{2}}},
\end{equation}
\end{small}
\begin{equation}
\Delta p = \left[\frac{1}{2}\left( {1 - \frac{{\omega {\xi ^{'}}}}{{\Omega {\xi ^5}}}} \right)
\sqrt {\frac{{{\xi ^2} - {\xi ^{ - 2}}}}{{{\xi ^2} - {\xi ^{'}}^{\rm{2}}}}}  + \frac{\omega }{{2\Omega {\xi ^6}}}\right]^{\frac{1}{2}}.
\end{equation}
Near the critical point, the critical behavior gives
\begin{equation}
  \begin{array}{l}
\Delta x \propto f\left( \lambda  \right){\left| \epsilon \right|^{ - 1/4}}, \quad
\Delta p \propto {f^{ - 1}}\left( \lambda  \right){\left| \epsilon \right|^{1/4}}.
\end{array}
\end{equation}
For $p$-type superradiant phase,  $\Delta x$ and $\Delta p$ are obtained
via the ground state ${\psi _2}\left( x,\alpha_{2} \right) $.
\begin{equation}
\Delta x = {\left[ {\frac{1}{2}\left( {1 - \frac{{\omega \xi }}{{\Omega {\xi ^{'}}^5}}} \right)\sqrt {\frac{{{\xi ^{'}}^2 - {\xi ^{'}}^{ - 2}}}{{{\xi ^{'}}^2 - {\xi ^2}}}}  + \frac{\omega }{{2\Omega {\xi ^{'}}^6}}} \right]^{\frac{1}{2}}},
\end{equation}
\begin{sequation}
\Delta p = \left[\frac{1}{2}\left( {1 - \frac{{\omega \xi }}{{\Omega {\xi ^{'}}^5}}} \right)\sqrt {\frac{{{\xi ^{'}}^2 - {\xi ^2}}}{{{\xi ^{'}}^2 - {\xi ^{'}}^{ - 2}}}} + \frac{{\omega {\xi ^2}}}{{2\Omega {\xi ^{'}}^4}} \\
-  {\frac{\xi }{{2{\xi ^{'}}^3}} + \frac{\xi }{{2{\xi ^{'}}^7}}}  \right]^{\frac{1}{2}}.
\end{sequation}
Near the critical point, the critical behavior gives
\begin{equation}
  \begin{array}{l}
\Delta x \propto f^{-1}\left( \lambda  \right){\left| \epsilon \right|^{ 1/4}}, \quad
\Delta p \propto {f}\left( \lambda  \right){\left| \epsilon \right|^{-1/4}}.
\end{array}
\end{equation}
In the $x$-type superradiant phase, $\Delta x$ acts as the diverging length scale, while
in the $p$-type  superradiant phase, $\Delta p$ is the diverging length scale.
According to Eq.~(\ref{Healing_Relaxation_Scaling}), the diverging length scale gives the critical exponent $v=1/4$ and the dynamical critical exponent $z=2$.

\begin{figure*}[htbp]
\centering
\includegraphics[scale=0.88]{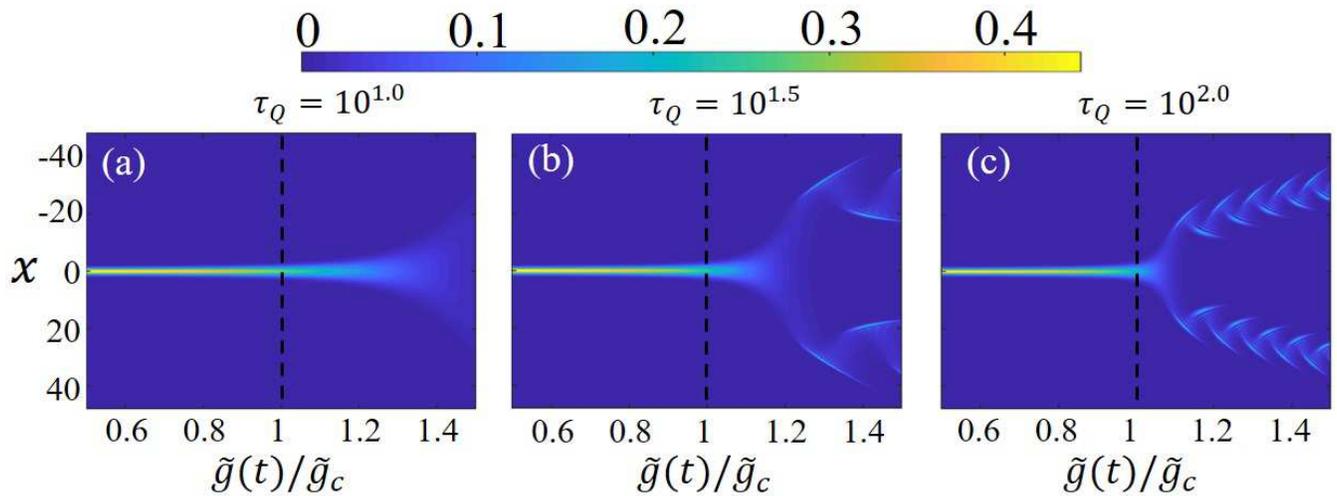}
\caption{(Color online) The time-evolution of density distribution for different quench times $\tau_{Q}$.
When the system is quenched from normal to superradiant phases at a finite $\tau_{Q}$, due to the vanishing excitation gap at the critical point, the system can't adiabatically across the critical point and the wave-packet spreads due to the appearance of excitations.
 The parameters are chosen as $L=96$, $\omega=1$, $\Omega=1000$, $\lambda=1$ and $\tau_{Q}={10^{1.0}, 10^{1.5}, 10^{2.0}}$.}
\label{State_Evolution_1}
\end{figure*}

\subsection{Numerical scalings}

\begin{figure}[htbp]
\centering
\includegraphics[scale=0.44]{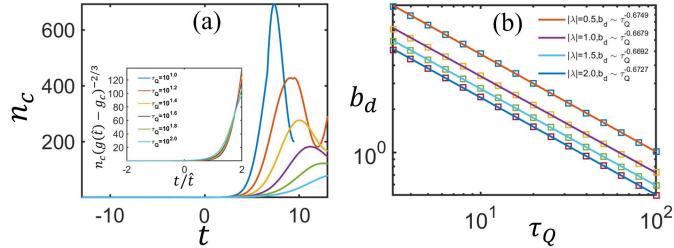}
\caption{(Color online) (a) The time evolution of the phonon number $n_{c}$ for different quench times, the inset shows the universal behavior in the rescaled time.
(b) The universal scaling of the phase transition delay $b_{d}$ with respect to the quench time $\tau_{Q}$ for different  $\lambda$.}
\label{bd_vs_time_tauq_1}
\end{figure}

Below we show how to numerically extract the Kibble-Zurek scalings from the non-equilibrium dynamics.
We perform the numerical simulations based on the Hamiltonian~(\ref{Hamiltonian1}).
To study the non-equilibrium dynamics, we prepare the initial ground state deeply in the normal phase,  in order to drive the system cross the superradiant phase transition, the coupling strength $\tilde{g}$ is linearly quenched according to
\begin{equation}\label{quench_way}
  \tilde{g}(t)=\tilde{g}_{c}\left(1-t/\tau_{Q}\right),
\end{equation}
where $\tilde{g}_{c}$ is the critical point, and  $\tau_{Q}$ is the quench time.
The typical total density distributions for different quench times are shown in Fig.~\ref{State_Evolution_1}.
When the system is quenched at a fast rate (which respond to small $\tau_{Q}$), the state may remain the information of the normal phase even in the deep superradiant region, see Fig.~\ref{State_Evolution_1}(a).
However, the state evolves more adiabatic as the quench time becomes larger, in which the oscillation amplitude of the state becomes smaller, see Fig.~\ref{State_Evolution_1}(c).
 When $\tau_{Q}\to\infty$, the quench dynamic returns to the equilibrium case, see the Fig.~\ref{GS_1}(a).
Base on aforementioned description, the phonon number $n_{c}=\langle \omega (p^{2}+x^{2})/2 \rangle$ serves as the order parameter.
When the coupling strength $\tilde{g}$ is quenched across the phase transition $\tilde{g}_{c}$, the time-evolution of $n_{c}$ are shown in Fig.~\ref{bd_vs_time_tauq_1}(a).
In the case of equilibrium phase transition, the phonon number becomes non-zeros when the system sweeps through the critical point.
However,  in the case of the non-equilibrium dynamics,  the phonon number delays to increase until the system crosses the freeze time $\hat{t}$,  where the state restarts to evolve.

To study the phase transition delay $b_{d}$,  we define the $b_{d}$ as
\begin{equation}\label{Phase_transition_delay}
  b_{d}\sim\left|\epsilon\right|\sim\left|\tilde{g}(\hat{t})-\tilde{g}_{c}\right|\sim\tau_{Q}^{-\frac{1}{1+vz}},
\end{equation}
where $\hat{t}$ is the freeze time.
In our calculation, $\hat{t}$ is determined when the phonon number $n_{c}$ reaches at fixed value $n_{c}^{fix}$.
According to the KZM,  the instantaneous state freezes at $-\hat{t}$ but with global phase evolution during the impulse region.
Hence, the order parameter remains zero in the first adiabatic region and the impulse region.
When the system crosses over the freeze time $\hat{t}$, the instantaneous state restarts to evolve again, but the states at this moment are no longer the eigenstates of the Hamiltonian.
Therefore, the order parameter becomes nonzero after the freeze time $\hat{t}$.
We determine the freeze time $\hat{t}$ when the phonon number $n_{c}$ satisfies $n_{c}^{fix}=5$.
In Fig.~\ref{bd_vs_time_tauq_1}(b), we show the universal scaling of the the phase transition delay $b_{d}$ with respect to the quench time $\tau_{Q}$,  the numerical scalings  for different $\lambda$ are well consistent with the analytical result $b_{d}\sim\tau_{Q}^{-\frac{1}{1+vz}}\sim\tau_{Q}^{-2/3}$.

\begin{figure}[htbp]
\centering
\includegraphics[scale=0.44]{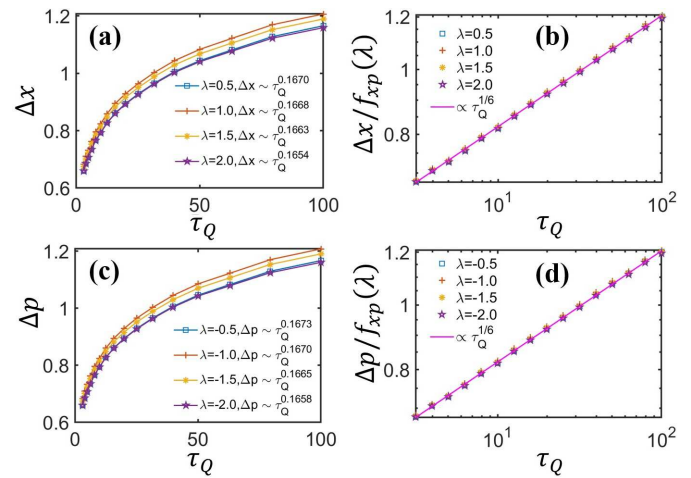}
\caption{(Color online) (a,c) Universal scalings of the diverging length scale $\Delta x$ and $\Delta p$ with respect to the quench time $\tau_{Q}$ for different  $\lambda$.
(b,d) Universal scalings of the  rescaled diverging length scale in the logarithmic coordinate.}
\label{Deltaxp_vs_tauq}
\end{figure}

Here, we treat the position variance $\Delta x$ (or the momentum variance $\Delta p$) as the diverging length scale when $\lambda>0$(or $\lambda<0$).
In Fig.~\ref{Deltaxp_vs_tauq}(a,b), we demonstrate the universal scaling of the $\Delta x$ with respect to the quench time $\tau_{Q}$.
The power laws are well agree with the analytical result $\Delta x \sim \tau_{Q}^{\frac{v}{1+vz}}\sim\tau_{Q}^{1/6}$.
In the case of $\lambda<0$, the momentum variance $\Delta p$ serves as the diverging length scale.
Similarly, $\Delta p$ shows universal scaling with respect to the quench time $\tau_{Q}$ as shown in Fig.\ref{Deltaxp_vs_tauq}(c,d), the power law is well consistent with the analytical result $\Delta p \sim \tau_{Q}^{\frac{v}{1+vz}}\sim\tau_{Q}^{1/6}$.
Combing the scalings of the phase transition delay and the diverging length scale, we finally give the numerical critical exponents of anisotropic QRM for different $\lambda$ in TABLE.~\ref{Critical_Exponent_Table}.

\begin{table}
\normalsize
\begin{center}
\begin{tabular}{|l|l|l|l|l|}
\hline
  Anisotropic ratio $\lambda$ & -0.5 & -1.0 & -1.5 & -2  \\ \hline

z(Critical exponent) & 1.994 & 1.989 & 2.017 & 2.009  \\ \hline

$\nu$(Critical exponent) & 0.2511 & 0.2501 & 0.2507 & 0.2486  \\ \hline

  Anisotropic ratio $\lambda$ & 0.5 & 1.0 & 1.5 & 2  \\ \hline

  z(Critical exponent) & 1.998 & 1.991& 2.019 & 2.013 \\ \hline

$\nu$(Critical exponent) & 0.2506 & 0.2497 & 0.2504 & 0.2480 \\ \hline
\end{tabular}
\caption{The numerical critical exponents(z,$\nu$) of the anisotropic QRM for different $\lambda$, which are well consistent with the analytical ones $(z=2,\nu=1/4)$. }
\label{Critical_Exponent_Table}
\end{center}
\end{table}

\section{Summary and Discussions}

In summary, we have investigated the non-equilibrium dynamics across a normal-to-superradiant phase transition in the anisotropic QRM.
Through performing the SW transformation, the Hamiltonian can be diagonalized and so that the ground states and their excitation gaps can be analytically obtained.
By analyzing the excitation gap and the diverging length scale, we give the critical exponents ($z=2,\nu=1/4$).
Meanwhile, we also simulate the real-time slow dynamics across the normal-to-superradiant phase transition.
To extract the critical exponents, we study the phase transition delay and diverging length scale near the critical point, which show universal scalings with respect to the quench time.
By introducing position and momentum operators, we clearly show the spontaneous symmetry breaking in the anisotropic QRM, which manifests the total density distribution spontaneously varies from a single-peak to double-peak shape.
Moreover, we reveal that the anisotropic QRM shares the same universal class (i.e. the identical critical exponents) in spite of the anisotropic ratio.

It is possible to realize the QRM in the ultrastrong coupling regime and the deep strong coupling regime via superconducting circuits~\cite{Langford2017,Braumuller2017,Leroux2018}, cold atoms~\cite{Felicetti2017} and trapped ions~\cite{Pedernales2015,Lv2018}.
The realization of the anisotropic QRM is more challengeable, some attempts have been proposed via quantum well~\cite{Schliemann2003,Wang2016}, circuit QED systems~\cite{Baksic2014,Yang2017}, superconducting flux qubits~\cite{WangMY2018}.
The ratio between the atomic transition frequency $\Omega$ and the phonon field frequency $\omega$ can be tuned by adjusting  the frequency detuning of the time-dependent magnetic fields, the qubits frequency and the LC oscillator frequency.
The coupling interaction strength $g$ and the anisotropic ratio $\lambda$ can be tuned by adjusting the phase of the time-dependent magnetic fields.

\begin{acknowledgments}
This work is supported by the Key-Area Research and Development Program of GuangDong Province under Grants No. 2019B030330001, the National Natural Science Foundation of China (NNSFC) under Grants No. 11874434 and No. 11574405, and the Science and Technology Program of Guangzhou (China) under Grants No. 201904020024.
\end{acknowledgments}


\bibliography{S_KZM_Reference2}

\end{document}